\documentclass[aps,reprint,twocolumn]{revtex4}
\usepackage{amsmath}
\usepackage{graphicx}

\draft 

\begin{document}


\title{Formation Mechanism of Bound States in Graphene Point Contacts} 



\author{Hai-Yao Deng$^1$}
\author{Katsunori Wakabayashi$^1$}
\email[Corresponding author:]{WAKABAYASHI.Katsunori@nims.go.jp}
\author{Chi-Hang Lam$^2$}
\email[Corresponding author:]{C.H.Lam@polyu.edu.hk}
\affiliation{$^1$International Center for Materials Nanoarchitechtonics (WPI-MANA),
National Institute for Materials Science (NIMS), Namiki 1-1, Tsukuba
305-0044, Japan}  
\affiliation{$^2$Department of Applied Physics, The Hong Kong Polytechnic University, Hung Hom, Hong Kong}


\date{\today}

\begin{abstract}
Electronic localization in narrow graphene constrictions are
 theoretically studied, and it is found that long-lived ($\sim 1~\mbox{ns}$) 
 quasi-bound states (QBSs) can exist in a class of 
 ultra-short graphene quantum point contacts (QPCs). These QBSs are shown
 to originate from the dispersionless edge states that are
 characteristic of the electronic structure of generically terminated
 graphene in which pseudo time-reversal symmetry is broken. The QBSs can be
 regarded as interface states confined between two graphene samples and
 their properties can be modified by changing the sizes of the QPC and
 the interface geometry. In the presence of bearded sites, these QBS can
 be converted into bound states. Experimental consequences and potential
 applications are discussed.    
\end{abstract}


\maketitle 

\section{Introduction}
\label{section:1}
Quantum point contacts (QPCs), which are narrow constrictions connecting two wider
samples, constitute fundamental building blocks of miniaturized devices
such as quantum dots and qubits~\cite{Houten1996,Zhang2011}. Being open
systems, QPCs are usually incapable of supporting atomistically small
quasi-bound states (QBSs)~\cite{Houten1992,Thomas2010}. However, if
they exist, QBSs can radically affect the properties of a system. For
example, they might trap electrons and produce local magnetic
moments~\cite{Iqbal2013,Bauer2013,Yakimenko2013,Yoon2009,Rejec2006,Hirose2003,Thomas1996,Thomas1998},
which can cause spin-dependent transpsort through a QPC.  

\begin{figure}
\begin{center}
\includegraphics[width=0.45\textwidth]{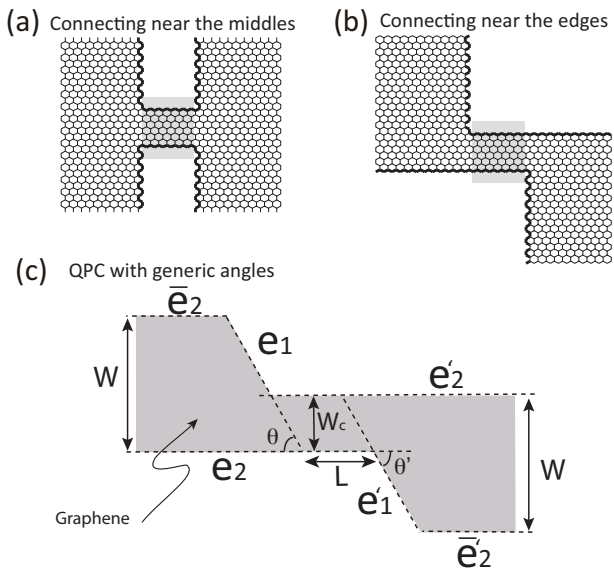}
\end{center}
\caption{(a) Schematic of the QPC connecting samples near the middle. (b)
 Schematic of the QPC connecting samples near the edge, where edge states
 can play an important role in electron transmission. (c) Schematic of the
 generic graphene QPCs connecting samples near the edge. Both samples
 infinitely extend along the edges $e_2$ and $e'_2$. The angles $\theta$
 and  $\theta'$ can be taken to be the same without loss of generality.}    
\label{figure:f1v}
\end{figure}

Graphene, which is a one-atom-thick carbon sheet, has attracted
tremendous attention in the past decade owing to its novel physical
properties and potential applications for future electronic
devices~\cite{Geim2007,White2007a}. Nanostructures made of graphene can be
patterned using lithography technique~\cite{Tapaszto2008}. Graphene QPCs
have been fabricated and extensively
studied~\cite{Girdhar2013,Guttinger2012,Ozyilmaz2007,Terres2011,Han2010,Stampfer2008,Darancet2009,Todd2008}.  
A shortest-possible QPC, which is made of a single hexagon and makes an
aperture for electrons, has been theoretically
examined~\cite{Darancet2009}, and typical wave diffraction patterns
were predicted. To date, all graphene QPCs investigated have been designed
to connect the middle of samples, as sketched in
FIG.~\ref{figure:f1v}(a), and no signatures of electron localization were
found in the ballistic limit.    

In this paper, we systematically study a different type of graphene QPC,
 where two graphene samples are connected near the edges as shown in
FIG.~\ref{figure:f1v}(b). In these QPCs, the edge states, which appear
on zigzag-shaped graphene edges at zero
energy~\cite{Fujita1996a,Fujita1996b,Enoki2013}, are shown to dominate
the electronic transport properties. For half graphene plane with a
perfect zigzag edge, the edge states are non-bonding and
are located on only one of the two sublattices. It has been shown that the
edge states are crucial in determining the magnetic and transport properties of
nanostructured graphene
systems~\cite{Rycerz2007,Wakabayashi2007,White2010,White2008,White2007,Wakabayashi2002,Fujita1996b,Wakabayashi2012,Karimi2012,Akhmerov2008,Deng2013}.  

In conventional graphene QPCs, the edge states have negligible effects
because they are far from the QPC. However, we show that electrons can
be localized in QPCs as depicted in
FIG.~\ref{figure:f1v}(b), i.e., where edge states located on different
sublattices are coupled, resulting in the formation of QBSs. These QBSs can live up to
$\tau_{QBS}\sim\mbox{ns}$ for sufficiently large samples and their wave
functions spread over only a few lattice constants. Their lifetimes can
be tuned by changing the geometry of the QPC and the size of the sample,
whereas their energies $\varepsilon_{QBS}$ are found to be insensitive
to sample dimensions. These QBS may be used as few-level quantum
dots and artificial atoms.   

We organize the paper as follows. In Section~\ref{section:2}, we
classify the edge-connected graphene QPCs into three classes and give a
brief overview of the results. In Section~\ref{section:3}, we describe
the formation mechanism of QBSs using the Green's function approach. In
Section~\ref{section:4}, we apply the theory to an example QPC, where
analytical results are obtained and compared to numerical calculations.
Finally, in Section~\ref{section:6}, we discuss some experimental
signatures and potential applications.   

\begin{figure*}
\begin{center}
\includegraphics[width=0.95\textwidth]{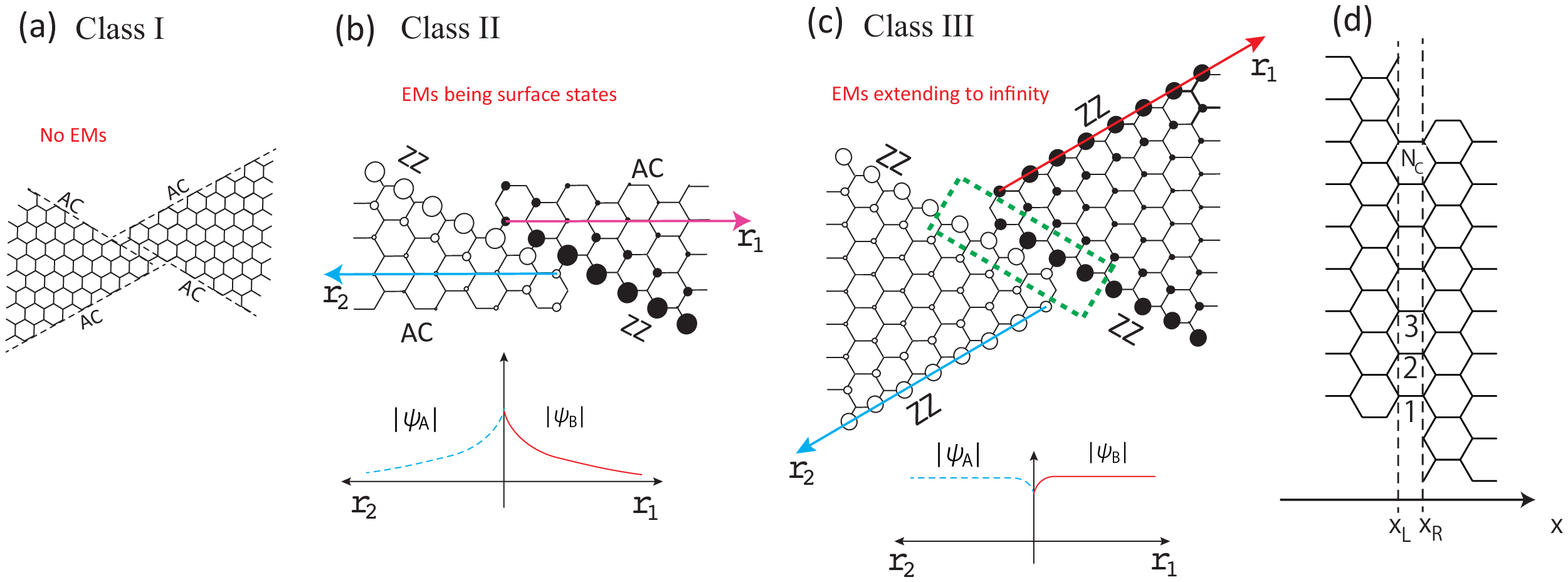}
\end{center}
\caption{Classification of QPCs connecting samples near the edges. (a)
 Example of Class I, which has all AC edges and no edge
 state. QPCs in this class are similar to QPCs connecting near the
 middle of samples. (b) Example of Class II, for which has the interface edges
 are ZZ but the extending edges are AC. In this class the edge
 states exponentially decay from the interface and are localized near
 the interfaces, i.e. bound states. Here the white (black) circles
 indicate the charge density of edge states on the A (B) sublattice
 sites. The schematic charge densities of edge state are for isolated
 samples~\cite{Shimomura2011}. (c) Example of Class III, for which the
 extending edges are ZZ whereas the interfacing edges can be either ZZ
 (as in the example shown here) or AC (as in the example shown in
 FIG.~\ref{figure:f2v}). In this class, edge states also exist. However,
 they are not bound states as indicated by the schematic wave
 functions of semi-infinite isolated samples (bottom panel).
 Nevertheless, bound states emerge when QPCs are present. (d) Enlarged
 view of QPC [indicated by the dashed rectangle in panel (c)]. The QPC contains
 $N_c$ connecting bonds.}  
\label{figure:f2vv}
\end{figure*}

\section{Overview}
\label{section:2}
Figure~\ref{figure:f1v}(c) schematically shows the QPC that connects two
graphene samples of the same width $W$ near their edges. This QPC can be
taken as an aperture for electron waves. The length and width of the QPC
are $L$ and $W_c$, respectively. Each graphene sample is geometrically
confined by three edges, which are denoted by $e_1$, $e_2$, and
$\bar{e}_2$ ($e'_1$, $e'_2$, and $\bar{e'}_2$) for the left-hand 
(right-hand) sample. We have assumed that $e_2$ ($e'_2$) is parallel with 
$\bar{e}_2$ ($\bar{e'}_2$). This condition is not necessary but facilitates
the analysis. The interface edges (i.e., $e_1$ and $e'_1$) are assumed
to be parallel for a smooth joint. Furthermore, we presume each edge to be
either a perfect armchair (AC) or zigzag (ZZ) (i.e., the angles $\theta$
and $\theta'$ are integers of $\frac{\pi}{6}$). According to the edge
structure, we can classify the QPCs into three classes of configurations
as shown in FIG.~\ref{figure:f2vv}. The details of each class are
described below.   

(Class I) All edges in a QPC are AC, as shown in
FIG.~\ref{figure:f2vv}(a). Because there are no edge states in this class,
the resulting QPCs resemble conventional QPCs and are therefore not
further addressed in this paper.   

(Class II) The edges $e_1$ and $e'_1$ are ZZ, whereas the edges $e_2$
and $e'_2$ are AC. An example is given in FIG.~\ref{figure:f2vv}(b). In
this class edge states appear on $e_1$ and $e'_1$. However, these edge
states are no more than surface states, whose wave functions are bound
to $e_1$ ($e'_1$), i.e., the wave function exponentially decays away from the QPC as
schematically shown in FIG.~\ref{figure:f2vv}(b). Because the edge states
do not extend along $e_2$ ($e'_2$), they do not directly participate in
electronic transport. Therefore, this class of QPCs simply
admixes the surface states bound on $e_1$ with those on $e'_1$,
yielding (quasi-) bound states.   

(Class III) Our main interest lies in this third class. The interface
edges $e_1$ and $e'_1$ can be either AC or ZZ, while the edges $e_2$ and
$e'_2$ (and hence $\bar{e}_2$ and $\bar{e'}_2$) are ZZ. There are
two bunches of edge states located on different sublattices; these edge states
extend along $e_2$ ($e'_2$) and $\bar{e}_2$ ($\bar{e'}_2$).
In FIG.~\ref{figure:f2vv}(c), we show an example of
all ZZ edges. A different geometry is shown in
FIG.~\ref{figure:f2v}, where the interface edges are AC.  
 
The QBSs are formed because of the non-bonding nature of the edge states
of the left and right samples (i.e. edges $e_2$ and
$e_2^\prime$). The QBS energy $\pm\varepsilon_{QBS}$ and lifetime
$\tau_{QBS}$ are sensitive to the ratio $W_c/W$. QBSs have a long lifetime
only for $W_c\ll\frac{W}{2}$. 
The quantity $\tau_{QBS}$ rapidly increases with
$W$ according to a power law;  $\tau_{QBS}\sim
\left(\frac{W}{W_0}\right)^3$ps,
whereas $\varepsilon_{QBS}$ is insensitive to $W$.
Here $W_0$ is a length scale, which is around
$86a\approx 21\mbox{nm}$ for $\theta=\theta'=\frac{\pi}{2}$. In general,
QBS causes resonant scattering which leads to a resonance peak in the
conductance $g$ of the QPC. For large $W$, $g$ takes on a symmetric
Breit-Wigner form,   
\begin{equation}
g\approx g_m\frac{\Gamma^2_{QBS}}{(\varepsilon_{F}-\varepsilon_{QBS})^2+\Gamma^2_{QBS}}, 
\label{g}
\end{equation}
where $g_m\sim 1$ in units of $g_0=\frac{2e^2}{h}$, $\varepsilon_F$ is
the Fermi level and $\Gamma_{QBS}=\frac{\hbar}{\tau_{QBS}}$ denotes the
level-broadening parameter. For finite $W$, the background contribution
leads to an asymmetric lineshape for $g$, i.e., Fano resonances.     

\section{Theory: $T$-Matrix Formalism}
\label{section:3}
\subsection{Model}
In this section, we analyze the electronic properties of graphene QPCs
using transition matrix ($T$-matrix) formalism based on the
nearest-neighbor tight-binding model~\cite{Economou1979}. We refer to
the geometry shown in FIG.~\ref{figure:f2vv}(c) for clarity. The
enlarged view of the QPC is displayed in FIG.~\ref{figure:f2vv}(d),
where the two graphene samples are connected via $N_c$ connecting
bonds. Each bond has a left- and right-hand end lying on $x=x_L$ and
$x=x_R$, respectively. The end sites on $x=x_L$ ($x_R$) are labeled 
$i_L$ ($i_R$), where the index $i_L$ ($i_R$) runs over 
$1,2,\cdots, N_c$. The graphene edge along $x=x_L$ ($x_R$) corresponds
to edge e$_1$ (e$_1^\prime$) in FIG.~\ref{figure:f1v}(c).   

The total Hamiltonian of the system can be written as $H=H_L+H_R+V$,
where $H_{L}$ ($H_R$) describes the left (right) isolated graphene
sample while $V$ stands for the QPC. Explicitly, we have
\begin{equation}
V=\sum^{N_c}_{i_L=1}\sum^{N_c}_{i_R=1}
\gamma_{i_L,i_R}
\left(|i_L\rangle\langle i_R|+ {\rm h.c.}\right),
\end{equation}
where $\gamma_{i_L,i_R}=-\gamma_0\delta_{i_L,i_R}$ with $\gamma_0\approx
2.7\mbox{eV}$ and $\delta_{i,j}$ is the Kronecker's function. We
introduce the bare Green's function,
$G_0(\varepsilon)=(\varepsilon+i0_+-H_0)^{-1}$, with $H_0=H_L+H_R$. For
later use, we resolve the diagonal elements of $G_0$ as follows:  
\begin{equation}
G^{i,i}_0(\varepsilon)=\sum_{\mu}|\psi_{i;\mu}|^2(\varepsilon+i0_+-\varepsilon_{\mu})^{-1},
 \quad \psi_{i;\mu}=\langle i|\mu\rangle, 
\label{G0}
\end{equation}   
 where $|\mu\rangle$ are the eigenstates of $H_0$, i.e.,
 $H_0|\mu\rangle=\varepsilon_{\mu}|\mu\rangle$. Now the $T$-matrix can
 be defined as 
\begin{equation}
T(\varepsilon)=(1-VG_0)^{-1}V=V+VG_0V+(VG_0)^2V+\cdots
\label{T}
\end{equation}
where $0_+$ denots an infinitesimal positive number. In principle, this
matrix captures all physical effects arising from scattering at the
QPC. The QBS can be found by searching for the poles of 
$T(\varepsilon)$.  

\begin{figure*}
\begin{center}
\includegraphics[width=0.9\textwidth]{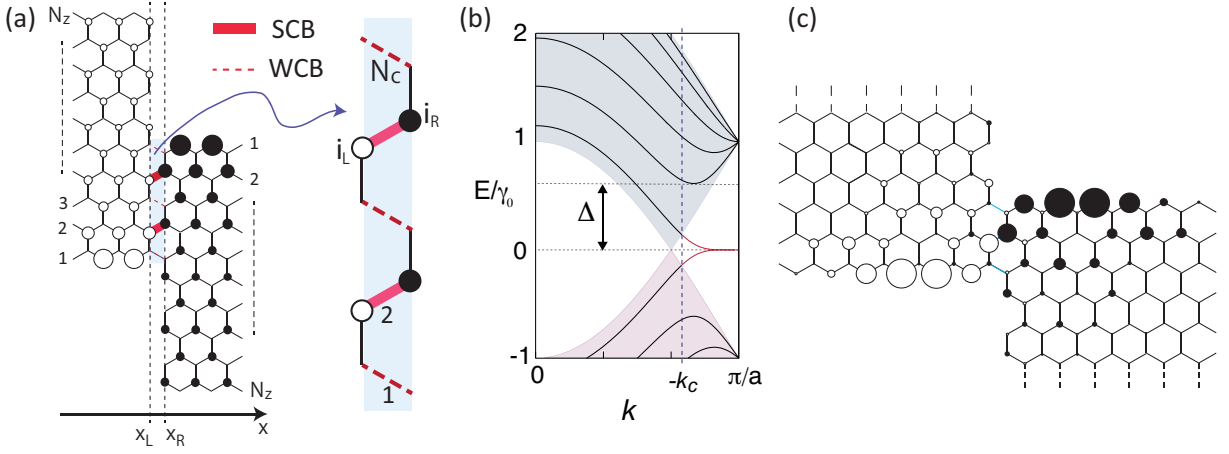}
\end{center}
\caption{(a) Class-III QPC with rectangular corners with $N_c=5$ connecting bonds. In
 this QPC, the edge states have amplitude on sublattice A (indicated by
 white circle) for the left-hand graphene sample, whereas they have
 amplitude on sublattice B
 (indicated black circle) for the right-hand graphene sample. The SCBs and WCBs
 alternate with each other. (b) Energy spectrum for a semi-infinite
 ribbon consisting of $N_z=6$ zigzag chains. Shaded areas indicate the spectrum of bulk
 graphene. Edge states appear in $k_c< k\le \pi/a$ (red segments). The
 black lines indicate the extended states. (c) Magnitude of the QBS wave
 function $\langle j|QBS\rangle$ in the vicinity of a SCB assuming a
 symmetric boundary condition, which can be observed in STM. The calculation was done for $N_z=50$ and
 $\varepsilon_{QBS}\approx 0.04$.}    
\label{figure:f2v}
\end{figure*}

For further analysis, let us closely inspect the bonding character at
the interface. We note that the $N_c$ connecting bonds fall into two
categories: strongly connecting bonds (SCBs) and weakly connecting
bonds (WCBs). Introducing the bare local density of states (LDOS) on
atomic site $i$ at energy $\varepsilon$ as
$\rho_0(\varepsilon,i)=-\frac{1}{\pi}Im[G^{i,i}_0(\varepsilon)]$, 
a SCB is then defined to have nonvanishing $\rho_0(0,i)$ on both $i=i_L$ and
$i=i_R$ where $i_L$ and $i_R=i_L$ denote the sites belonging to this
bond. Similarly, a WCB is defined to lack this property. In the QPC
shown in FIG.~\ref{figure:f2vv}(c), all connecting bonds are
SCBs. However, in the QPC shown in FIG.~\ref{figure:f2v}, which also
belongs to Class III, the SCBs and WCBs alternate with each other.   

\subsection{Energy and Lifetime of QBS}
In general, it is a formidable task to evaluate $T$ exactly. Here we
use two approximations. First, we neglect all inter-bond
transitions (i.e., $T_{i_R,i_L}\propto \delta_{i_R,i_L}$). This is
reasonable, because these transitions are higher-order processes in $G_0$
compared with intra-bond transitions. Second, we assume
$G^{i,i}_0\approx 0$ at low energies if $i$ belongs to a WCB, which can
be justified in the limit $\frac{W}{2}\gg W_c$. Then, we can easily
derive that $T_{i_R,i_L}(\varepsilon)\approx -\gamma_0\delta_{i_R,i_L}$
for WCBs and that  
\begin{equation}
T_{i_R,i_L}(\varepsilon)\approx\frac{-\gamma_0\delta_{i_R,i_L}}{1-
\gamma^2_0G^{i_L,i_L}_0(\varepsilon)G^{i_R,i_R}_0(\varepsilon)},  
\label{Tf}
\end{equation}
for SCBs~\cite{note0}. Basically, this expression describes the physical processes in
which an electron travels back and forth between the sites $i_L$ and
$i_R$, in analogy with back-and-forth bounces experienced by an
electron sandwiched between two potential
barriers~\cite{Thomas2010}. Note that such processes increase returning
probability and are responsible for the formation of QBSs.   

The energy and broadening of QBS can be determined by seeking the poles
of Eq.~(\ref{Tf}). Rewriting
$\gamma^2_0G^{i_L,i_L}_0(\varepsilon)G^{i_R,i_R}_0(\varepsilon)=R^{i_L,i_R}(\varepsilon)+i\cdot
I^{i_L,i_R}(\varepsilon)$, the poles
$z_{QBS}=\varepsilon_{QBS}+i\Gamma_{QBS}$ can be obtained via 
\begin{equation}
1-R^{i_L,i_R}(z_{QBS})-i\cdot I^{i_L,i_R}(z_{QBS})\approx 0.
\label{pole}
\end{equation}
As will be shown later, in the large $W$ limit, we have 
\begin{equation}
G^{i_\tau,i_\tau}_0\approx\frac{1}{\varepsilon}\cdot B_\tau -\frac{i}{\gamma_0}\cdot C_\tau,
\label{GEM}
\end{equation}
where $\tau=$L or R. $B_\tau$ and $C_\tau$ are real values, which vary
from bond to bond. Later, we will see that $C_\tau$ decreases to zero 
with increasing $W$ following a power law whereas $B_\tau$ approaches a constant
$B_{\tau,\infty}$. Upon substitution, we immediately find 
\begin{eqnarray}
\begin{pmatrix}
\varepsilon_{QBS}\\ 
\Gamma_{QBS}
\end{pmatrix}
&\approx&
\frac{-\gamma_0}{2(1+C_LC_R)}\nonumber\\
&\cdot&
\begin{pmatrix}
\pm\sqrt{4B_LB_R-(B_LC_R-B_RC_L)^2} \\
B_LC_R+C_LB_R
\end{pmatrix}
\label{result}
\end{eqnarray}
From this it follows that, (1) there is a pair of QBS with each SCB,
whose energies are symmetric about zero; (2) the energy of the QBS is
not sensitive to the sample width, i.e.,
$\varepsilon_{QBS}\approx\pm\gamma_0\sqrt{B_LB_R}\approx \pm\varepsilon_{\infty}=\pm\gamma_0\sqrt{B_{L,\infty}B_{R,\infty}}$; (3) the QBS has a broadening, $\Gamma_{QBS}$, which shrinks rapidly with $W$, as explained below.   

It proves useful to rewrite Eq.~(\ref{Tf}) as 
\begin{equation}
T_{i_L=l,i_R=l}(\varepsilon)=-\frac{\gamma_0\varepsilon^2}{(\varepsilon-z_{QBS,l})(\varepsilon+z^*_{QBS,l})},
\label{Tt}
\end{equation}   
where we have used Eq.~(\ref{GEM}) and $l$ indicates the $l$-th SCB. The maximum of
$|T_{i_L=l,i_R=l}(\varepsilon)|$ occurs when
$\varepsilon=\varepsilon_{QBS,l}$, for which we have 
\begin{equation}
T_{i_L=l,i_R=l}(\varepsilon_{QBS,l})=-i\gamma_0\frac{\varepsilon_{QBS,l}}{2\Gamma_{QBS,l}}\gg T_{i_L\neq l,i_R\neq l}(\varepsilon_{QBS,l}).
\label{Tm}
\end{equation}
Evidently, the $l$-th SCB predominates for $\varepsilon \sim \varepsilon_{QBS,l}$. 

Now we establish Eq.~(\ref{GEM}). For this purpose, we decompose
Eq.~(\ref{G0}) into its real and imaginary parts: 
\begin{equation}
G^{i_\tau,i_\tau}_0(\varepsilon)=\sum'_{\mu}|\psi_{i_\tau;\mu}|^2(\varepsilon-\varepsilon_{\mu})^{-1}-i\pi\rho_0(\varepsilon,i_\tau) 
\label{Go}
\end{equation}
where $i_\tau$ belongs to a SCB,
$\rho_0(\varepsilon,i_{\tau})=\sum_{\mu}|\psi_{i_\tau;\mu}|^2\delta(\varepsilon-\varepsilon_{\mu})$
is the aforementioned LDOS on site $i_{\tau}$ and $\delta$ denotes the
Dirac $\delta$ function while the prime indicates the principal value of
the sum (which can be turned into an integral). The $\mu$ denotes either
an extended state or an edge state. In the sum, the contributions from
the extended states are of the order $W^{-3}$\cite{note2} and can then
be neglected for large $W$, whereas the contributions from the edge
states are roughly independent of $W$. Therefore, considering that edge
states have zero energies at large $W$, we arrive at the $G^{i,i}_0$ as
given in Eq.~(\ref{GEM}), with the coefficients given by 
\begin{eqnarray}
B_\tau &=& \sum_{\mu=\mbox{edge states}}|\psi_{i_\tau;\mu}|^2,\nonumber\\ 
C_\tau &=& -\pi\gamma_0\rho_0(\varepsilon_{QBS},i_\tau).
\label{Bb}
\end{eqnarray}
From $B_{\tau}$ one determines $\varepsilon_{QBS}$. The quantity
$C_{\tau}$ is found by comparing Eqs.~(\ref{GEM}) and
$(\ref{Go})$. Equations~(\ref{Tf}), (\ref{GEM}) and (\ref{Bb}) constitute the
foundation of the present theory. They are applicable to all QPC
configurations exemplified in FIG.~\ref{figure:f2vv} in the limit
$\frac{W}{2}\gg W_c$. In the next section, we discuss prototypical
examples.  

In the limit $W\rightarrow\infty$, the edge states approach those of two
isolated half infinite graphene planes. Accordingly, the quantity $B_{\tau}$ tends to a
constant $B_{\tau,\infty}$ given by Eq.~(\ref{Bb}) with the edge
states of two half planes. Simultaneously, $C_{\tau}$ tends to zero,
since $\rho_0(\varepsilon_{QBS},i_{\tau})$ comes from only extended
states whose wave functions vanish on $i_{\tau}$ as $W/W_c\rightarrow 
\infty$.  

\subsection{Wave Functions of QBS}
To derive the QBS wave function, we shall consider the QBS
associated with the $l$-th SCB, for which $i_L=i_R=l$. 
In the conventional scattering theory~\cite{Economou1979}, the state vector is given by
$|QBS,l\rangle=|\psi_0\rangle+G_0T|\psi_0\rangle$, where
$H_0|\psi_0\rangle=\varepsilon_{QBS,l}|\psi_0\rangle$. We have
explicitly included the index $l$ to indicate the QBS in question. To
specify the QBS wave function, $\Psi_{l}(j)\equiv \langle
j|QBS,l\rangle$, where $j$ denotes an arbitrary site in the entire
system, we have to impose boundary conditions on $\langle
j|\psi_0\rangle$. Two types of boundary conditions are considered here.

(Type I) This type assumes an open system and is appropriate for
studying transport properties. The $\Psi_l(j)$ is supposed to describe
an electron wave incident from the  
left-hand sample, tunneling through the QPC and partailly transmitted to
the right-hand sample. 
The $\langle j|\psi_0\rangle$ describes the
superposition of the incident wave and the totally reflected wave. We then
find  
\begin{eqnarray}
\Psi_l(j)&\approx& \langle j|\psi_0\rangle\nonumber\\&+&T_{i_L,i_L}(\varepsilon_{QBS,l})G^{j,i_L}_0(\varepsilon_{QBS,l})\langle i_L|\psi_0\rangle\nonumber\\ &+& T_{i_R,i_L}(\varepsilon_{QBS,l})G^{j,i_R}_0(\varepsilon_{QBS,l})\langle i_L|\psi_0\rangle, 
\label{wf}
\end{eqnarray}
where $i_L=i_R=l$. The third term describes the transmitted wave. Note that we have kept only the $l$-th bond, which is reasonable according to Eq.~(\ref{Tm}). 

(Type II) This type assumes a closed system, in which no current flows
from left to right. It is suitable for describing scanning tunneling
microscopy (STM). 
We then find 
\begin{eqnarray}
\Psi_l(j)&\approx& \langle j|\psi_0\rangle + T_{i_L,i_R}(\varepsilon_{QBS,l})\nonumber\\ &\times&[G^{j,i_L}_0(\varepsilon_{QBS,l}) \cdot\langle i_R|\psi_0\rangle \nonumber\\ 
&+&G^{j,i_R}_0(\varepsilon_{QBS,l})\cdot \langle i_L|\psi_0\rangle].
\label{wfq1}
\end{eqnarray}
where $i_L=i_R=l$. We have neglected the terms headed by $T_{i_L,i_L}$ and $T_{i_R,i_R}$, which are smaller than the retained terms~\cite{note0}. An example of $\Psi_l(j)$ is mapped in FIG.~\ref{figure:f2v}(c), where we see that $\Psi_l(j)$ extends over only a few lattice constants in space.  

\section{Example: $\theta=\theta'=\frac{\pi}{2}$}
\label{section:4}
Let us illustrate the above theory for the rectangular-corner
configuration shown in FIG.~\ref{figure:f2v}. In this case, isolated
graphene samples are semi-infinite ribbons. Their wave functions can be
obtained from those for infinite ribbons, for which analytical solutions
have been established~\cite{Wakabayashi2012}. Thus, the coefficients
$B_\tau$ and $C_\tau$ can be analytically obtained. For convenience, we
express the width $W$ in terms of the number of total zigzag chains
$N_z$ as $W=\frac{a}{2\sqrt{3}}(3N_z-2)$.   

\subsection{Calculation of $\varepsilon_{QBS}$ and $\Gamma_{QBS}$}
The goal is to find the coefficients $B_\tau$ and $C_\tau$. For this
purpose, we need $\psi_{i;\mu}$, the wave function of mode
$|\mu\rangle$, which can be represented as a superposition of two
counter-propagating waves related by time reversal symmetry appropriate
to an ideal zigzag graphene ribbon. They can be easily constructed 
so we simply quote the results here:
\begin{equation}
\begin{pmatrix}
\psi_{i_L;\mu}\\
\psi_{i_R;\mu}
\end{pmatrix}
=\sqrt{\frac{2a}{L}}e^{ika}\sin\left(\frac{ka}{2}\right)
\begin{pmatrix}
-\Phi_{\mu}(i_L)\\
\Phi_{\mu}(i_R)
\end{pmatrix}
\label{psi}
\end{equation}
In Eq.~(\ref{psi}), $k\in[0,\pi/a]$ and $L$ denotes the circumference of a
virtual zigzag graphene tube used to discretize the values of
$k$. $\Phi$ is the transverse component of the wave fuction. Note that
$\mu$ is shorthand for a composite index, $(k,u,s)$, where
$u=1,\cdots,N_z$ counts the subbands and $s=\pm$ is the particle-hole
label, as sketched in FIG.~\ref{figure:f2v}(b).   

Specific to each subband, there is a quantum number $p_u$, which is real
at any $k$ for $u<N_z$~\cite{Wakabayashi2012}. However, for $u=N_z$, $p$ is real only if
$k\in[0,k_c]$ and it can be written as $p=\pi+i\eta$ if
$k\in(k_c,\frac{\pi}{a}]$. For this particular subband, in the large
$N_z$ limit, whereby $k_c=\frac{2\pi}{3a}$, it holds as a good
approximation that $\varepsilon_s(k)\approx 0$ and~\cite{note4}  
\begin{eqnarray}
\begin{pmatrix}
\Phi_{k,s}(i_L)\\
\Phi_{k,s}(i_R)
\end{pmatrix}
\approx \sqrt{\frac{1-g^2_k}{2}}
\begin{pmatrix}
g^{i_L-1}_k \\
g^{N_c-i_R}_k
\end{pmatrix}
\label{phie}
\end{eqnarray}
for $k>\frac{2\pi}{3a}$. Here $g_k=2\cos(ka/2)$. For $k<\frac{2\pi}{3a}$, in the same
limit, we instead have $\varepsilon_s(k)\approx s(g_k-1)$ and  
\begin{eqnarray}
\begin{pmatrix}
\Phi_{k,s}(i_L)\\
\Phi_{k,s}(i_R)
\end{pmatrix}
\approx \sqrt{\frac{1}{N_z}}
\begin{pmatrix}
\sin(pi_L)\\
\sin [p(N_c+1-i_R)
\end{pmatrix}
\label{phiee}
\end{eqnarray}
where $p\approx(1-\frac{1}{N_z})\pi$ and we have dropped the subscript
$u$ for this subband. Note that, as shown in Fig.~\ref{figure:f2v}(b),
this subband is the only one available within the energy window
$[-\Delta,\Delta]$, where $\Delta\approx 4\gamma_0\cos(\frac{N_z-1}{2N_z-1}\pi)$.   

To evaluate $C_\tau$, we may presume that the QBS lies inside
the single-channel energy window (i.e.,
$\varepsilon_{QBS}\in[-\Delta,\Delta]$). Then, the only contribution to
$\rho_0$, i.e. $C_\tau$, comes from the $u=N_z$ subband with
$k\in[0,\frac{2\pi}{3a}]$, because these are the only states available in
that energy window. This assumption, whose validity can be examined by
consistency check, implies that the QBS life time is essentially set by
the dispersing segment of the lowest subband. By using Eqs.~(\ref{psi}) and
(\ref{phiee}), we obtain  
\begin{equation}
\begin{pmatrix}
C_L \\
C_R
\end{pmatrix}
\approx A\cdot\frac{\pi^2}{N^3_z}
\begin{pmatrix}
i^2_L \\
(N_c+1-i_R)^2
\end{pmatrix}
\label{Im}
\end{equation}
with $A=-\frac{2a}{L}\cdot\pi\cdot\sum_{0\leq
k<\frac{2\pi}{3a}}\sin^2\left(\frac{ka}{2}\right)\delta[\varepsilon^2_{QBS}-(g_k-1)^2]$. Transforming
it into an integral, we find
$A=-\frac{\sin(x_0)}{2|\varepsilon_{QBS}|}$, with
$x_0\in(0,\frac{\pi}{3})$ given by
$|\varepsilon_{QBS}|=2\cos(x_R)-1$. The quadratic dependences on $i_L$ and
$N_c$ are notable in Eq.~(\ref{Im}), which explains why long-lived QBSs
only form when $N_c$ is small.   

We proceed to estimate $B_\tau$. At energies near zero the primary
contributions stem from the edge states. Actually, since
$p_u\approx\frac{u-1}{N_z}\pi$ for any extended state
[Fig.~\ref{figure:f2v}(b)] of any subband in the large $N_z$ limit, the 
total contributions from the low energy sector, i.e., including those
with $u\sim N_z$, are of the order $\sim N^{-1}_z \sum_{u\sim
N_z}|\sin(\frac{u-1}{N_z}\pi)|^2\approx
N^{-3}_z$\cite{note2}. Nonetheless, the contribution from the edge
states is of order unity, as indicated in Eq.~(\ref{phie}). Thus, when
$N_z$ is large, the edge states dominate. If we neglect the dispersion
of these states, which is reasonable for large $N_z$, we immediately
confirm Eq.~(\ref{Bb}). Using Eqs.~(\ref{psi}) and (\ref{phie}), we find  
\begin{eqnarray}
&\quad&\lim_{N_z\rightarrow\infty}
\begin{pmatrix}
B_L \\
B_R
\end{pmatrix}
=
\begin{pmatrix}
B_{L,\infty}\\
B_{R,\infty}
\end{pmatrix}
\\&=&\frac{2a}{L}\sum_{\frac{2\pi}{3a}<k<\frac{\pi}{a}}\sin^2(ka/2)(1-g^2_k)\nonumber
\begin{pmatrix}
g^{2(i_L-1)}_k \\
g^{2(N_c-i_R)}_k
\end{pmatrix}
\end{eqnarray}
which quickly diminish as $N_c$ or $i_L$ increases. 

Note that the maximum values of $B_{L,\infty}$ and $B_{R,\infty}$
occur at $(N_c=2,i_L=i_R=2)$, in which case one finds $B_{L,\infty}\approx 0.04$
and $B_{R,\infty}\approx 0.21$, leading to $\varepsilon_{\infty}\approx
0.09\gamma_0$. Therefore, for most ribbons of interest, we indeed
have $\varepsilon_{\infty}\in[-\Delta,\Delta]$, which is consistent
with our initial assumption~\cite{note3}. Another case of special
interest is $N_c=3$, for which we find $B_{L,\infty}=B_{R,\infty}\approx
0.04$, yielding $\varepsilon_{\infty}\approx 0.04\gamma_0$. We then see
that the $\varepsilon_{\infty}$ depends strongly on $N_c$. The
parameters for other interesting cases have also been calculated and
are tabulated in Table I.    

 \begin{table*}
 \caption{\label{table:1} Theoretically evaluated parameters for certain
  SCBs $(N_c,i_L)$ (obviously $i_R=i_L$ for a given bond) in the configuration shown in
  FIG.~\ref{figure:f2v}. From Eqs.~(8) and (19), we see that
  $\Gamma_{QBS}N^3_z$ (given in the last column) is independent of
  $N_z$. All energies are in units of $\gamma_0$.} 
 \begin{tabular}{c c c c c c c c}
\hline\hline
$(N_c,i_L)$ & $B_{L,\infty}$ & $B_{R,\infty}$ & $\varepsilon_{\infty}$ & $|A|$ & $C_L\cdot N^3_z$ & $C_R\cdot N^3_z$ & $\Gamma_{QBS}\cdot N^3_z$ \\ [0.5ex] 
\hline
(2,2) & 0.04 & 0.21 & 0.09 & 9.7 & 386.7 & 96.7 & 42.5 \\
(3,2) & 0.04 & 0.04 & 0.04 & 21.8& 870 & 870 & 34.8 \\ 
(4,2) & 0.04 & 0.017& 0.026& 33.5& 1338& 3011& 71.5 \\ 
(4,4) & 0.009& 0.21 & 0.044& 19.7& 3149& 196.8& 331.6 \\ 
\hline
 \end{tabular}
 \end{table*}

\subsection{Spatial Profile of QBS}
To visualize the QBS in real space, we calculated the wave function
of the QBS according to Eq.~(\ref{wfq1}) for symmetric boundary
condition. We neglected the first term in these equations, so the
spatial profile of the QBS is completely determined by $G^{j,i_L}_0$ and
$G^{j,i_R}_0$, which can be easily evaluated numerically using the
resolution of Eq.~(\ref{G0}). In FIG.~\ref{figure:f2v}(c), we show the
results for the symmetric configuration $(N_c=3,i_L=2)$. As expected, the
amplitudes are concentrated about the SCB, spreading over a few lattice
constant. This distribution can be observed in STM (see Section~\ref{section:6}).
It is worth noting that the SCB resembles a molecular junction
between the graphene samples.   

\begin{widetext}
\subsection{Conductance}
\label{section:5}
Ultra-narrow QPCs usually strongly reflect incident electron waves, as
would be anticipated from diffraction theory in the sub-wavelength
regime\cite{Bethe1944}. However, such reflections can be suppressed due
to resonant tunneling from QBSs. In what follows, we calculate the
conductance $g$ of the QPC shown in FIG.\ref{figure:f2v}(a) and derive Eq.~(\ref{g}).  

We focus on the single-channel regime, i.e., $\varepsilon\in(0,\Delta)$,
where the modes can each be labeled by just a wave number. We use $k$
and $q$ to denote the wave numbers for the left- and right-hand samples, respectively. Following standard tunneling theory\cite{Bardeen1961}, we obtain the conductance
at zero temperature as 
\begin{equation}
g = L^2v^{-2}_F|\langle q=k_F|T|k=k_F\rangle|^2, 
\label{g2}
\end{equation}
where $v_F=\left(\frac{d\varepsilon_k}{dk}\right)_{k=k_F}\approx-a\gamma_0\sin(k_Fa/2)$ 
and $k_F\in(0,\frac{2\pi}{3a})$ denote the Fermi velocity (in units of $\hbar=1$) and Fermi wave number, respectively. By using Eq.~(\ref{Tf}), we find 
\begin{equation}
\langle q|T|k\rangle\approx
-\gamma_0\sum_{WCB}\psi^*_{i_R,q}\psi_{i_L,k}+\sum_{SCB}\psi^*_{i_R,q}\psi_{i_L,k}T_{i_R,i_L},\nonumber
\end{equation}
Substituting this in Eq.(\ref{g2}), we find  $g=g_w+g_s+g_{ws}$, where $g_w$ ($g_s$) involves only WCBs (SCBs) whereas $g_{ws}$ involves both SCBs and WCBs. Explicitly, we have
\begin{eqnarray}
\begin{pmatrix}
g_w \\
g_{ws} \\
g_s
\end{pmatrix}
=L^2v^{-2}_F\cdot
\begin{pmatrix}
\gamma^2_0|\sum_{l=WCB}\psi^*_{i_R=l;k_F}\psi_{i_L=l;k_F}|^2 \\
-\gamma_0\sum_{l=SCB}\sum_{l'=WCB}
\psi^*_{i_R=l',k_F} 
\psi_{i_L=l',k_F}
\psi^*_{i_R=l,k_F}
\psi_{i_L=l,k_F}
T_{i_R=l,i_L=l}(\varepsilon_F) \\
|\sum_{l=SCB}\psi^*_{i_R=l;k_F}\psi_{i_L=l;k_F}T_{i_R=l,i_L=l}(\varepsilon_F)|^2
\end{pmatrix}
\end{eqnarray}
For $\varepsilon_F$ near the energy of a QBS, these terms scale with $N_z$ as follows,
\begin{equation}
g_w \sim N^{-6}_z, \quad g_{ws} \sim N^{-3}_z, \quad g_s \sim N^0_z. 
\end{equation} 
which can be shown on the basis of two observations. Firstly, from Eqs.~(\ref{psi}) and (\ref{phiee}) it follows that $\psi_{i_{\tau};k} \sim N^{-3/2}_z$ in the limit $N_c\ll\frac{N_z}{2}$. Secondly, from Eqs.~(\ref{Tt}) and (\ref{Tm}) it follows that $T_{i_R=l,i_L=l} \sim -\gamma_0 N^3_z$ if $\varepsilon_{l,QBS} \sim \varepsilon_F$ or $T_{i_R=l,i_L=l}\approx -\gamma_0$ otherwise. From this, we see that in the large-$N_z$ limit the dominant contribution to $g_{ws}$ and $g_s$ stems from the SCB whose energy is the closest to $\varepsilon_F$. Now the scaling becomes clear: in $g_w$, the wave functions contribute the $N^{-6}_z$ factor; in $g_{ws}$, this factor is raised by $N^3_z$ due to the $T$-matrix element, which contributes a $N^3_z$ factor; in $g_s$, two $T$-matrix elements appear and contribute $N^6_z$, which exactly cancels the $N^{-6}_z$ from the wave functions. 

The above analysis shows that, for $\varepsilon_F$ close to the energy of a QBS, the $g_w$ and $g_{ws}$ can be neglected for large $N_z$. Thus, we find
\begin{equation}
g\approx g_s = L^2v^{-2}_F|\psi^*_{i_R=l;k_F}\psi_{i_L=l;k_F}T_{i_R=l,i_L=l}(\varepsilon_F)|^2
\end{equation}
where $\varepsilon_F$ is near $\varepsilon_{QBS,l}$. By Eqs.~(\ref{Tt}) and (\ref{phiee}), this expression can be reduced to Eq.~(\ref{g}), with   
\begin{eqnarray}
g_m = 4\sin^2(k_la/2)|\Phi^*_{k_F}(i_R=l)\Phi_{k_F}(i_L=l)|^2\left(\frac{\varepsilon_{QBS,l}}{\Gamma_{QBS,l}}\right)^2 \sim N^0_z
\end{eqnarray}
where $k_l\approx\frac{2\pi}{3a}$ is given by $\varepsilon_{k_l}=\varepsilon_{QBS,l}\sim 0$. With the parameters given in Table I, it is easy to see that $g_m\approx 1$. We emphasize that Eq.~(\ref{g}) gives a good description only when $N_c\ll\frac{N_z}{2}$. Otherwise, additional contributions from WCBs and overlaps between adjacent QBS make the $g$ asymmetric and
more similar to a Fano resonance.  
\end{widetext}

\subsection{Numerical Calculations}
To verify the above results, we have performed numerical calculations
based on the Landauer formalism and mode-matching method. Details of
the scheme will be presented elsewhere.  

\begin{figure}
\begin{center}
\includegraphics[width=0.45\textwidth]{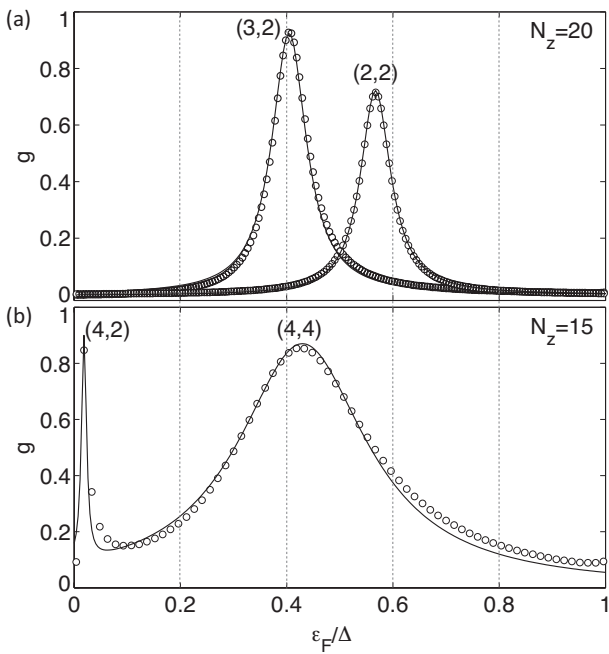}
\end{center}
\caption{Theory versus numerical calculations: energy dependence of
 conductance $g$ for the QPC shown in FIG.~\ref{figure:f2v}. Circles
 represent numerical calculations while solid lines indicate fitting
 according to Eq.~(1). The resonances are labeled $(N_c,i_L)$, in the
 same manner as in Table I. In panel (a), only one SCB exists, whereas
 in panel (c) there two SCBs exist. Each SCB leads to a peak in $g$.}    
\label{figure:F3}
\end{figure}

The calculated conductances for wide ribbons with $N_c=2,3,4$ are
presented as circles in FIG.~\ref{figure:F3}. For $N_c=2,3$, one resonance
peak is observed in the entire single-channel regime, as shown in 
FIG.~\ref{figure:F3}(a). Such peaks are interpreted as consequences of QBS, whose energy
and lifetime set the position and half-width of the peaks. As seen in
FIG.~\ref{figure:F3}(a), the line-shape of each peak can be well captured by
Eq.~(\ref{g}), which is a Lorentzian (solid curves). For $N_c=4$, as
shown in FIG.~\ref{figure:F3}(b), two peaks are observed. The lower energy
peak is very sharp whereas the higher energy peak is much 
broader. These peaks are identified with QBSs belonging to the two
SCBs, $(N_c=4,i_L=2)$ and $(N_c=4,i_L=4)$. The line-shape can
be described by a superposition of two Lorentzians, as noted in
FIG.~\ref{figure:F3}(b).       

\begin{figure}
\begin{center}
\includegraphics[width=0.4\textwidth]{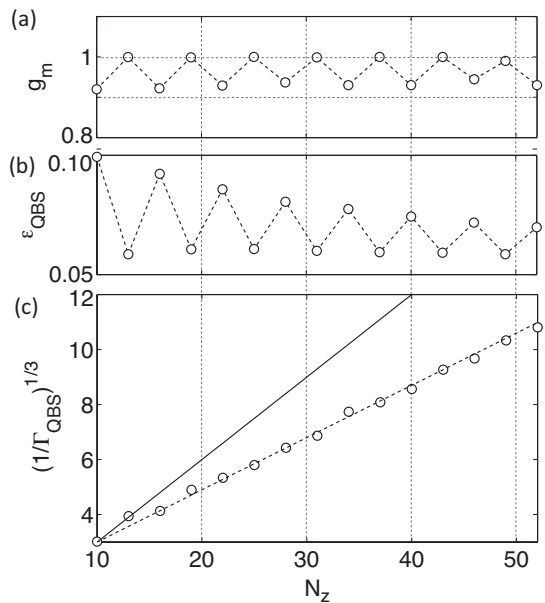}
\end{center}
\caption{Theory versus numerical calculations: $N_z$ dependences of (a)
 peak conductance, (b) QBS energy and (c) QBS broadening parameter (both
 in units of $\gamma_0$). The numerical calculations were done for the resonance
 $(3,2)$ seen in FIG.~\ref{figure:F3}(a). Our theory predicts roughly
 constant $g_m$, convergence of $\varepsilon_{QBS}$ to
 $\varepsilon_{\infty}$ and $\Gamma_{QBS}\sim N^{-3}_z$, all of which
 are consistent with numerical calculations. In panel (c), the solid curve has
 a slope of $\sim 0.3$ while the dashed curve has a slope of $\sim
 0.2$. The dashed line is to guide the eye. Solid line shows results of
 theory and circles show results of numerical calculations.}
\label{figure:F2}
\end{figure}

In FIG.~\ref{figure:F2} we examine the $N_z$ dependences of three
quantities: $g_m$, $\varepsilon_{QBS}$ and $\Gamma_{QBS}$ in FIGs.~\ref{figure:F2}(a),
(b) and (c), respectively. According to the theory, we expect (1) the
$g_m$ to be roughly independent of $N_z$, (2) the $\varepsilon_{QBS}$ to
converge to $\varepsilon_{\infty}$ and (3) the $\Gamma_{QBS}$ to
decrease as $N^{-3}_z$. All these features are borne out in numerical calculations,
as evident in FIGs.~\ref{figure:F2} (a), (b) and (c). We notice a weak
dependence of $g_m$ and $\varepsilon_{QBS}$ on the parity of $N_z$. This
odd-even effect gradually disappears when $N_z$ increases beyond
$\sim250$, which may be understood by observing that the edge state dispersion 
can be written as $\varepsilon_{s}(k)\approx 2s(-1)^{N_z+1}[1+2\cos(ka)]g^{N_z}_k$ in 
the large $N_z$ limit. The factor $(-1)^{N_z+1}$ may be the origin of such effects. 
For sufficiently large $N_z$ ($N_z>\sim250$), we have $\varepsilon_s(k)<\varepsilon_{\infty}$ for 
all $k\in(\frac{2\pi}{3a},\frac{\pi}{a}]$. Eq.~(\ref{GEM}) still holds and we 
have $[B_L,B_R]\approx \frac{2a}{L}\sum_{k,s}\sin^2(\frac{ka}{2})(1-g^2_k)(1-\varepsilon_{s}(k)/\varepsilon_{\infty})^{-1}[g^{2(i_L-1)}_k,g^{2(N_c-i_R)}_k]$, which does not display any parity 
effect. However, for $N_z$ not that large ($N_z<\sim250$), then $\varepsilon_{\infty}$ will 
cut $\varepsilon_s(k)$ and parity effects can appear. Nonetheless, this case is 
not amenable to analytical expressions and will not be further discussed.    

\subsection{Role of Bearded Sites}
\begin{figure}
\begin{center}
\includegraphics[width=0.4\textwidth]{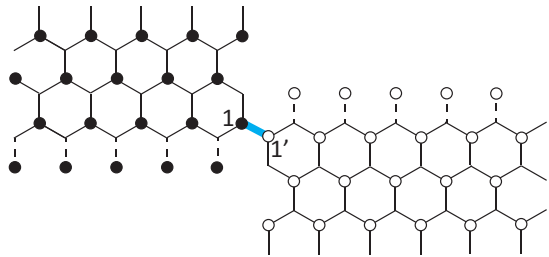}
\end{center}
\caption{Bearded sites, represented by the circles attached to the
 dashed bonds, turn a WCB into a SCB. A pair of real bound states form
 on the bond (see text). Geometrically, the SCB (blue bond) shown here
 is similar to the SCB $(N_c=3,m=2)$ without bearded sites.}   
\label{figure:F4}
\end{figure}

Here we discuss the effect of bearded sites and show that they could
lead to genuine bound states (i.e., vanishing $\Gamma_{QBS}$). For
simplicity, we consider the rectangular-corner QPC with $N_c=1$ 
(FIG.~\ref{figure:F4}).  In the absence of bearded sites, the connecting
bond would be a WCB. However, bearded sites transform an edge state
initially located on the A- (B-) sublattice to edge state located on the B- (A-)
sublattice~\cite{Wakabayashi2001}, and thus turn a WCB into a SCB. Then,
one can show that $C_\tau=0$ and then the level broadening vanishes,
implying a pair of genuine bound states on this bond. Actually, bearded
graphene is an insulator\cite{Wakabayashi2001} with a real band gap
separating the edge states from the extended states. Thus, the $\rho_0$
vanishes at the QBS energy, which is consistent with vanishing
$C_\tau$. The values of $B_\tau$ in the limit $N_z\rightarrow\infty$ are
expected to be similar to those of the SCB $(N_c=3,i_L=2)$ in the same
limit without bearded sites, because the wave functions in both cases are
the same [Eq.~(17)]. 

\section{Discussions}
\label{section:6}
An essential element in the QBS theory presented above relates to the
existence of non-bonding edge states at zero energy. Generically, such
states are originated from the breaking of pseudo time-reversal
symmetry, which is unique to the graphene lattice. Note that a perfect
zigzag edge is not necessary for them to appear. Indeed, they could
show up in graphene edges of almost any shape except for the perfect
armchair one (in which case the symmetry is respected), even in the
presence of a moderate external magnetic field~\cite{Gusynin2009}. The
observation makes our theory more widely applicable. 

Because the QBS extends over only a few lattice constants, the Coulomb
repulsion might be relatively strong. Assuming an on-site repulsion of
$10~\mbox{eV}$~\cite{Wehling2011}, the repulsion between two electrons in
QBS can be as large as $0.1~\mbox{eV}$, much bigger than that
in conventional semiconductor quantum dots ($\sim\mbox{meV}$). These
interactions serve to manipulate spins for possible applications in
spintronics and qubits. For sufficiently wide samples, the QBS lifetime
can be very long. As a result, charges may accumulate in the QPC and may
give rise to dynamical Coulomb blockade effects~\cite{Bulka2007}.  

The QBS may be visualized using STM,
which probes the dressed local density of states (LDOS)
directly~\cite{Tersoff1985}. One can show 
that in comparison with the bare LDOS, 
the dressed LDOS is enhanced by an amount 
\begin{math}
\delta\rho(\varepsilon_F,\vec{r})\sim
\frac{\Gamma_{QBS}}{(\varepsilon-\varepsilon_{QBS})^2+\Gamma^2_{QBS}}\cdot
f(\vec{r}), 
\end{math}
where $f(\vec{r})$ is a function that decays as the STM tip
moves away from a SCB by $|\vec{r}|$, as indicated in FIG.~\ref{figure:f2v}(c). The properties of the QBS can thus
be directly determined.  

The QBS can also have optical signatures. Specifically, we predict the
optical absorption to be enhanced at the frequency $\nu\approx
\frac{2\varepsilon_{\infty}}{h}$, which corresponds to the energy
required to excite an electron from the lower QBS at
$-\varepsilon_{\infty}$ to the upper one at $\varepsilon_{\infty}$. For
$(N_c=3,m=2)$, this gives $\nu\approx 48\mbox{THz}$ in the infrared
regime. Note that the QBS lifetime in this case is $\tau_{QBS}\approx
7\cdot\left(\frac{N_z}{100}\right)^3\mbox{ps}$, which can be much longer
than the optical oscillation period $\nu^{-1}$. Thus, the system may be
treated as an artificial two-level atom when dealing with its
interaction with light at frequencies near or above $\nu$.    

In addition, the QPCs can serve as the channel for a text-book
single-level (and single-electron in the presence of Coulomb
interactions) resonant tunneling transistor. Due to small level
broadening, sharp turn-on can be expected at low temperatures.  

\section{Conclusion}
\label{section:7}
In conclusion, we elucidated a mechanism for the formation of
atomic bound states in a type of graphene QPCs. These states arise
because of the zero energy edge states that are associated with the breaking of
pseudo time-reversal symmetry. Their energies have been shown to be
roughly independent of the sample dimensions. Finite level broadening
exists, which shrinks to zero following a power law as the sample width
increases. Because of the broadening, the states show up as Breit-Wigner
resonances in the conductance of the QPCs. Such resonances dominate the
electronic transport properties in the low energy regime.   

K. W. acknowledges the financial support by Grant-in-Aid for Scientific Research from MEXT and JSPS (Nos. 25107005, 23310083 and 20001006). C. H. L. thanks the support from HK PolyU through grant No. G-YM41.


\begin{thebibliography}{9}
\bibitem{note0}Employing the same approximations, we can derive from Eq.(4) that $T_{i_{\tau},i_{\tau}} \approx -\gamma_0G^{i_{\bar{\tau}},i_{\bar{\tau}}}_0T_{i_{\bar{\tau}},i_{\tau}}$, where $\tau=L,R$ and $\bar{\tau}=R,L$. We note that for energies close to zero, $G^{i_{\tau},i_{\tau}}_0\approx 0$ because of the particle-hole symmetry. Thus, $T_{i_{\tau},i_{\tau}}$ is much smaller than $T_{i_{\tau},i_{\bar{\tau}}}$.
\bibitem{note4} In Eq.~(\ref{phie}), the factor $2$ accounts for the fact
	that of the two degenerate sets of the edge modes obtained in the
	semi-infinite limit, only one contributes near either edge.
\bibitem{note3}We see that the $\Delta$ decreases with $N_z$. For our
	assumption to hold, we must have
	$\Delta>\varepsilon_{\infty}\sim 0.01$. Note that $\Delta\approx
	\frac{20}{2N_z-1}$, from which we infer that the widest possible
	ribbon should not exceed $N_z \sim 1000$, i.e., around
	$150\mbox{nm}$. If $N_z$ indeed exceeds this critical value, one
	can show that Eq.~(\ref{Im}) still holds but the coefficient $A$
	has to sum over more channels and subsequently, the
	$\tau_{\mbox{QBS}}$ scales as $N^2_z$ instead of $N^3_z$ (The
	additional factor accounts for the number of dispersive bands).
\end{thebibliography}
\end{document}